\documentclass[twocolumn,showpacs,preprintnumbers,amsmath,amssymb,prb]{revtex4}

\usepackage{graphicx}
\usepackage[usenames]{color}
\usepackage{dcolumn}
\usepackage{bm}

\begin{document}

\title{Transmission phase shift of phonon-assisted tunneling through a quantum dot}
\author{Juntao Song$^{1}$, Qing-feng Sun$^{1}$, Hua Jiang$^{1}$, and X. C. Xie$^{2,1}$}

\affiliation{$^1$Beijing National Lab for Condensed Matter Physics
and Institute of Physics, Chinese Academy of Sciences, Beijing
100080, China \\
$^2$Department of Physics, Oklahoma State University, Stillwater
Oklahoma, 74078 USA}

\date{\today}

\begin{abstract}
The influence of electron-phonon interaction on the transmission
phase shift of an electron passing through a quantum dot is
investigated by using the scattering theory. The transmission
phase versus the intra-dot level shows a serial of phonon-induced
dips. These dips are highly sensitive to electron-phonon
interaction strength $\lambda$, and they are much more pronounced
than phonon-assisted sub-peaks appeared in the conductance.
Phonon-induce dephasing is also studied, and the results show that
the dephasing probability $T_d$ monotonically increases with the
electron-phonon interaction strength $\lambda$. The dephasing
probability $T_d \propto\lambda^2$ for small $\lambda$ but
$T_d\propto \lambda$ at large $\lambda$.
\end{abstract}


\maketitle

\section{Introduction}

Electron transport through a mesoscopic system, e.g. a quantum dot
(QD), has been extensively investigated in the last two decades.
Due to the fact that the size of a mesoscopic device is within the
phase coherent length, the phase of a wave function plays a key
role on the electronic transport. So the transmission amplitude
$t=|t|e^{i\theta}$, which describes the electron tunneling through
a mesoscopic system, is a complex number. Its magnitude square
$|t|^2$ is the transmission probability which is observable in the
measurement of current or conductance. The transmission phase
$\theta$ describes the phase change when an electron tunnels
through a device. This phase $\theta$ is in general lost in the
measurement of current or conductance, so that it is difficult to
acquire in experiment. Using an AB interference ring device,
Yacoby {\sl et al.}~\cite{A. Yacoby} tried, for the first time, to
measure the transmission phase $\theta$ through a QD. Couple years
later, Schuster {\sl et al}\cite{R. Schuster} utilized an open
multi-terminal AB ring device to successfully measure the phase
$\theta$. Since then, to investigate the transmission phase has
generated a great deal of theoretical and experimental interest
with a fare amount of efforts focusing in this field. On the
experimental side, for example, Buks {\sl et al.}~\cite{E. Buks}
reported that controlled decoherence could be achieved in a device
with a QD that is capacitively coupled to a quantum point contact
in close vicinity. The phase evolution in the Kondo regime was
experimentally investigated a few years back,\cite{Yang Ji, Ulrich
Gerland} and was found to be highly sensitive to the onset of
Kondo correlation. Recently, Leturcq {\sl et al.}~\cite{R.
Leturcq} investigated the magnetic field symmetry and the phase
rigidity of the nonlinear conductance in a AB ring. On the other
hand, the success of these experiments generates a number of
theoretical studies. In early of 1980s, Buttiker found the phase
rigidity in a two-terminal AB ring device due to the time-reversal
symmetry and the current conservation.\cite{addref1} After the
experiment by Schuster {\sl et al.},\cite{R. Schuster} many
follow-up theoretical efforts focused on and tried to interpret
the measured results of the transmission phase $\theta$ through a
QD, in particular, the abrupt lapses of $\theta$ between two
adjacent resonances and the similar behavior of $\theta$ for all
resonant peaks.\cite{addref2,G. Hackenbroich} In addition, some
works have also studied the transmission phase in the Kondo
regime,\cite{Ulrich Gerland} or with the photon-assisted tunneling
process under a time-dependent external field,\cite{addref5} etc.

Another subject, the electron-phonon (e-ph) interaction in a
single-molecular QD, has also generated a great deal of interest
in recent years. The phonon-assisted tunneling peaks or steps have
been experimentally observed in various single-molecule transistor
systems.\cite{H. Park,B. J. Leroy,S. Sapmaz} Park {\sl et
al.}~\cite{H. Park} observed phonon-assisted tunneling sub-steps
in the I-V curves in a single-$C_{60}$ transistor device, and
those sub-steps are attributed to the coupling of electron and the
$C_{60}$-surface vibration mode. In another experiment by Leroy
{\sl et al}.,\cite{B. J. Leroy} the current and the conductance of
a suspended individual single-wall nanotube device are measured,
and the phonon-assisted sub-peaks on the two sides of the main
resonant peak are clearly visible in the differential conductance
versus gate voltage, which is due to the radial breathing phonon
mode. On the theoretical side, the influence of e-ph interaction
on the mesoscopic transport is also studied by several
groups.~\cite{J. T. Song,J. X. Zhu,N. S. Wingreen} Many
interesting results, e.g. the phonon-assisted sub-peaks, etc, are
first theoretically predicted, and then experimentally observed.

In this paper, we investigate the transport behavior of a
molecular QD system having an e-ph interaction by using the
scattering matrix method. We focus mainly on the transmission
phase of the phonon-assisted tunneling sub-peaks, as well as the
phonon-induced dephasing process. The results exhibit that the
transmission phase $\theta$ drops between two adjacent (sub)-peaks
and $\theta$ rises again near the position of sub-peaks. In
particular, the characteristic of phonon-assisted tunneling
process in the transmission phase is much more pronounced and
visible than these sub-peaks in the conductance. Afterwards we
discuss the dephasing ratio. While at zero temperature and at low
bias $V$ ($V<\omega_0$ with $\omega_0$ being the phonon
frequency), the electronic transport through the molecular QD is
completely coherent because the electron can not absorb or emit
phonons under this condition. However, if at non-zero temperature
or at a high bias $V_{bias}$ ($V_{bias}>\omega_0$), the dephasing
process occurs. In the limit of high bias $V_{bias}$ ($V_{bias}\gg
\omega_0$), the dephasing ratio goes as square of the e-ph
interaction strength $\lambda$ in the weak interaction region, but
it is linearly dependent on $\lambda$ in the strong interaction
region. In addition, we also consider an open AB ring device with
a molecular QD embedded in one of its arms, and find that it is
feasible to experimentally measure the influence of the e-ph
interaction through the transmission phase.

The rest of this paper is organized as follows. We introduce the
model and derive the formula of transmission amplitude in Sec. II.
In Sec. III, we present the numerical results and their
discussions. In Sec. IV, we study the phase measurement by using
an open AB ring device. Finally, a brief summary is presented in
Sec. IV. Some detailed derivation of the transmission amplitude is
given in Appendix.

\section{MODEL AND FORMULATION}

The system under consideration is a molecular QD coupled to left
and right leads in the presence of a local phonon mode, and it can
be described by the following Hamiltonian:
\begin{eqnarray}
H&=&H_0+H_1,
\end{eqnarray}
where
\begin{eqnarray}
H_0 & =&\sum_{\alpha,k}{\varepsilon_{\alpha k}c_{\alpha k}^\dagger
c_{\alpha k}}+\varepsilon_0{d^\dagger d}+\omega_0{b^\dagger b},\\
H_1 &= &\lambda (b^\dagger+ b){d^\dagger d}
+\sum_{\alpha,k}{[t_{\alpha k} c_{\alpha k}^\dagger d+H.c]}.
\end{eqnarray}
Here $c_{\alpha k}^\dagger (c_{\alpha k})$ and $d^\dagger (d)$ are
the electron creation (annihilation) operators in the lead
$\alpha=L,R$ and the QD, respectively. $b^\dagger (b)$ is the
phonon creation (annihilation) operator in the QD. Due to large
level spacing of the molecular QD, only one relevant quantum level
$\varepsilon_0$ is considered. The electron in the QD is coupled
to a single phonon mode $\omega_0$, and $\lambda$ and $t_{\alpha
k}$ describe the strength of the e-ph interaction and the coupling
between the QD and the leads, respectively.

In the following, we apply the S-matrix scattering formalism to
derive the transmission amplitude, the transmission phase and the
current. From Hamiltonian (1), the S matrix can be written
as:~\cite{J. R. Taylor,N. S. Wingreen}
\begin{eqnarray}
S&=&1-i\int^{\infty}_{-\infty}dt_1 e^{iH_0t_1}H_1e^{-iH_0t_1}e^{-\eta |t_1|}\nonumber\\
&&-i\int^{\infty}_{-\infty}\int^{\infty}_{-\infty}dt_1 dt_2
e^{iH_0t_2}H_1 \hat{G}_r(t_2-t_1)\nonumber\\
&&\times H_1e^{-iH_0t_1}e^{-\eta (|t_1|+|t_2|)},\ \ \eta\rightarrow
0^{+}
\end{eqnarray}
where the single-particle Green's function operator
${\hat{G}}_r(t)$ is, ${\hat{G}}_r(t)=-i\theta(t)e^{-iHt}$. By
using the $S$ matrix, the final state $|f>$ can be obtained from
the initial state $|i>$, with $ |f>=S\times|i>$. Considering an
initial state $|i>= |\varepsilon_i,n,L>$, which denotes an
electron with energy $\varepsilon_i$ in the left lead and n
phonons in the QD, the final state $|f>$ can be expressed as:
\begin{eqnarray}
|f>&=&S\times|i>=S\times|\varepsilon_i,n,L>\nonumber\\
&=&\sum_{m=-\infty}^{+\infty}[r_m(\varepsilon_f,\varepsilon_i)|\varepsilon_f,m,L>+t_m(\varepsilon_f,\varepsilon_i)|\varepsilon_f,m,R>]\nonumber\\
&=&\sum_{m=-\infty}^{+\infty}[r_m(\varepsilon_i)\delta(\varepsilon_i+n\omega_0-\varepsilon_f-m\omega_0)|\varepsilon_f,m,L>\nonumber\\
&&+t_m(\varepsilon_i)\delta(\varepsilon_i+n\omega_0-\varepsilon_f-m\omega_0)|\varepsilon_f,m,R>].\nonumber\\
\end{eqnarray}
where $t_m(\varepsilon)$ and $r_m(\varepsilon)$ are the
transmission amplitude and the reflection amplitude with
accompanying absorption or emission of $|m-n|$ phonons. At zero
temperature, the phonon number $n$ in the initial state $|i>$ must
be zero and then $t_m(\varepsilon_i)$ can be written as (the
detailed derivation is shown in the Appendix):
\begin{eqnarray}
t_m(\varepsilon_i)&=&\int d \varepsilon_f\ t_m(\varepsilon_f,\varepsilon_i)\nonumber\\
&=&-\frac{i}{\sqrt{m!}}{\Gamma}
e^{-\lambda ^2}\sum_{l=0}^ m\frac{(-1)^{m-l}m!}{l!(m-l)!}\nonumber\\
&&\times\sum^{\infty}_{n=0}\frac{\lambda^{2n+m}}{n!}\overline{G}^r(\varepsilon_i-n\omega_0-l\omega_0)\nonumber\\
\end{eqnarray}
Obviously, $t_0(\varepsilon)$ describes the amplitude of an
elastic tunneling process which is coherent. While
$t_{m}(\varepsilon)$ ($m\neq0$) is the amplitude of an inelastic
tunneling for emitting $m$ phonons. Due to emission of phonons,
thus, leaving a trace in the QD for the inelastic tunneling
process, an inelastically tunnelled electron loses its phase
coherence. So at zero temperature the transmission phase shift
through the QD is:~\cite{R. Schuster,G. Hackenbroich}
\begin{eqnarray}
\theta=arg\left\{t_0(0)\right\}.
\end{eqnarray}
From $t_m(\varepsilon)$, the total transmission probability
(including the coherent and the non-coherent parts) through the QD
is $T_{tot}(\varepsilon) = \sum_{m=0}^\infty
|t_m(\varepsilon)|^2$, and the transmission probability of the
non-coherent part is $T_d(\varepsilon)=\sum_{m=1}^\infty
|t_m(\varepsilon)|^2$.

\section{Numerical Results and Discussions}

In this section, we numerically study the transmission phase
$\theta$ and the dephasing ratio $T_d/T_{tot}$. In our numerical
calculations, the phonon frequency $\omega_0$ is set as the energy
unity ($\omega_0=1$). Notice that the main result of this paper,
Eq. (6), is obtained at zero temperature, and thus so are the
numerical results and their discussions. However, results at low
temperature should be similar to that at zero temperature. Fig. 1
shows the coherent transmission probability $|t_0|^2$ [namely,
$|t_0(0)|^2$] and the phase $\theta$, as a function of
renormalized level $\overline{\varepsilon}_0$, which can be tuned
by the gate voltage in an experiment. Notice that $|t_0|^2$ is
proportional to the linear conductance $G$ through the QD,
$G=(e^2/h) |t_0|^2$. Due to the e-ph interaction, several
interesting features are manifested. In addition to the main peak
related to the single level, new satellite sub-peaks appear in the
curve of $|t_0|^2$-$\overline{\varepsilon}_0$ at
$-\overline{\varepsilon}_0 =n\omega_0$ ($n=1,2,...$). The
sub-peaks only exist on the right hand side of the main peak, and
their heights increase with the increase of the e-ph coupling
strength $\lambda$. The sub-peak at
$\overline{\varepsilon}_0=-\omega_0$ corresponds to the
phonon-assisted tunneling process as shown in the inset of Fig.
1(a), in which an incident electron from the left lead first emits
a phonon and tunnels to the level $\overline{\varepsilon}_0$, and
subsequently reabsorbs a phonon and tunnels forward to the right
lead. Since this process does not leave a trace in the QD, it
maintains the phase coherence. Meanwhile at zero temperature,
there is no phonon in the QD in the initial state and the
absorption process can not occur, so that the satellite sub-peaks
only exist on the negative $\overline{\varepsilon}_0$ side [see
Fig. 1(a)].

Next, we study the transmission phase $\theta$ that exhibits a
non-monotonic behavior. Across the main resonant peak, $\theta$
continuously rises by a value of $\pi$. This result is consistent
with the previous theoretical and experimental findings.~\cite{R.
Schuster,addref2,G. Hackenbroich} Because of the e-ph interaction,
$\theta$ drops between the main peak and the next sub-peak or
between two adjacent sub-peaks, and rises again across a sub-peak,
such that a dip appears around $-\overline{\varepsilon}_0
=n\omega_0$. These dips are much more pronounced than those
sub-peaks in the transmission probability $|t_0|^2$. For example,
for $\lambda=0.8$ the 2-nd phonon-assisted sub-peak is so small
that it is hardly visible [see Fig. 1(a)], however even the 4-th
dip can be clearly seen [see Fig. 1(b)]. The sensitivity of the
transmission phase $\theta$ to the e-ph interaction provides a new
way to detect the strength of the e-ph interaction.

In the calculations above, the tunneling coupling strength
$\Gamma$ ($\Gamma=\Gamma_L=\Gamma_R$) between the leads and the QD
is set to be quite weak, $\Gamma=0.1 \ll \omega_0$. With an
increase of $\Gamma$, the phonon-assisted sub-peaks and the main
peak in the curve of $|t_0|^2$-$\overline{\varepsilon}_0$
gradually merge together and become indistinguishable, and the
dips in the curve of $\theta$-$\overline{\varepsilon}_0$ are also
gradually getting smaller (see Fig. 2). While $\Gamma \approx
\omega_0$ (e.g. $\Gamma =0.7$), all sub-peaks and all dips are
almost invisible. Consequently, in order to experimentally detect
the phonon-induced dips of the phase $\theta$ or the
phonon-assisted sub-peaks, the coupling strength $\Gamma$ should
be tuned to be less than $\omega_0/2$. In fact, the condition
$\Gamma<\omega_0/2$ is normally satisfied in the
experiments.\cite{B. J. Leroy}

Let us study the amplitude $t_m(\varepsilon)$ ($m=1,2,...$) of the
inelastic tunneling process. In this inelastic tunneling process,
an incident electron emits $m$ phonons while tunneling through the
QD. However it is prohibited when $V_{bias}< m \omega_0$. At small
bias case $V_{bias}<  \omega_0$ and at zero temperature, all
inelastic tunneling processes are prohibited and the tunneling
through the QD is coherent. On the other hand, with $V_{bias} >
\omega_0$, inelastic tunneling processes occur and the tunneling
through the QD is partly non-coherent. In the limit of larger bias
voltage, $V_{bias} \gg \omega_0$, the total dephasing transmission
probability $T_d(\varepsilon)$ is: $T_d(\varepsilon)
=\sum\limits_{m=1}^{\infty} |t_m|^2$. Fig. 3 shows the dephasing
transmission probability $T_d$ (or $T_d/T_{tot}$) versus
renormalized level $\overline{\varepsilon}_0$ and the e-ph
interaction strength $\lambda$. While without the e-ph interaction
(namely $\lambda=0$), no inelastic tunneling process happens and
thus $T_d=0$. When $\lambda\not=0$, the inelastic tunneling
process occurs and $T_d$ is no longer zero. A serial of peaks are
exhibited in the curve of $T_d$-$\overline{\varepsilon}_0$ and the
interval between the two adjacent peaks is $\omega_0$ [see Fig.
3(a)]. As is seen from Fig. 3(a), a higher peak must correspond to
a larger value of $\lambda$ (including the peak at
$\overline{\varepsilon}_0 =0$). This means that the dephasing
probability $T_d$ monotonously increases with the $\lambda$
regardless of the position of the renormalized level
$\overline{\varepsilon}_0$. Next, in Fig. 3(b) we show the
relative dephasing transmission probability $T_d/T_{tot}$ versus
the e-ph coupling strength $\lambda$ in the resonant tunneling
region (namely $\overline{\varepsilon}_0=0$). When $\lambda$ is
small ($\lambda<0.2\omega_0$), the relative dephasing transmission
probability $T_d/T_{tot}$ increases parabolically with $\lambda$,
but the dephasing probability $T_d/T_{tot}$ is found to increase
linearly with increasing $\lambda$ between the range of $1>\lambda
>0.4$. For large $\lambda$ case ($\lambda>1$), $T_d/T_{tot}>0.6$
and the dephasing inelastic tunneling processes dominate. In an
experiment the parameter $g=(\lambda/\omega_0)^2$ is generally in
the range from 0.1 to 1, ~\cite{S. Sapmaz} though some special
devices ~\cite{A. N. Pasupathy} show a big variable range of $g$.
In this $ \lambda$ region, the degree of dephasing is linearly
dependent on the e-ph coupling strength.

\section{the AB ring device}

In Sections II and III, we only consider a simple device
consisting of a QD coupled to two leads. However, in a real
experiment to measure the transmission phase $\theta$, the device
is an open AB ring with a QD embedded in one of the arms.\cite{R.
Schuster} Therefore, it is of experimental relevance to study the
open AB ring device in this section. A QD is embedded in one arm
of the ring, and the other arm is the reference arm with the
transmission amplitude $t_{ref}$. Due to openness of the open AB
ring device, the processes of multi-time circling around the ring
is negligible. Note that only elastic tunneling process
$t_0(\varepsilon_i)$ is in interference with the reference arm.
$T_{AB}(\varepsilon_f,\varepsilon_i)$, defined as the probability
that an electron of energy $\varepsilon_i$ incident from the left
lead will be transmitted with energy $\varepsilon_f$ into the
right lead, can therefore be written as:
\begin{eqnarray}
T_{AB}(\varepsilon_f,\varepsilon_i)=\sum_{m=0}^\infty\delta(\varepsilon_i-\varepsilon_f-m\omega_0)
|t_m(\varepsilon_i)+\delta_{m,0}\ e^{i\phi}t_{ref}|^2,\nonumber\\
\end{eqnarray}
where $\phi$ is the magnetic flux inside the ring. In the absence
of the reference arm (namely, $t_{ref}=0$),
$T_{AB}(\varepsilon_f,\varepsilon_i)$ is reduced to
$T(\varepsilon_f,\varepsilon_i) = \sum_{m=0}^\infty
\delta(\varepsilon_i-\varepsilon_f-m\omega_0)
|t_m(\varepsilon_i)|^2$, and this result is the same as that in
the work by Wingreen {\sl et al.}~\cite{N. S. Wingreen} Using the
transmission probability $T_{AB}(\varepsilon_f,\varepsilon_i)$,
the current flowing through the AB ring is:~\cite{N. S. Wingreen}
\begin{eqnarray}
J_{AB}&=&\frac{2e}{h}\int d\varepsilon_i\int d\varepsilon_f
T_{AB}(\varepsilon_f,\varepsilon_i)f_L(\varepsilon_i)[1-f_R(\varepsilon_f)]\nonumber\\
&&-\frac{2e}{h}\int d\varepsilon_i\int d\varepsilon_f
T_{AB}(\varepsilon_f,\varepsilon_i)f_R(\varepsilon_i)[1-f_L(\varepsilon_f)],\nonumber\\
\end{eqnarray}
where $f_L(\varepsilon_i) =f(\varepsilon_i -\mu_L)$ and
$f_R(\varepsilon_f) =f(\varepsilon_f -\mu_R)$ with the chemical
potential $\mu_{L(R)} = \pm e V_{bias}/2$ and $f(\varepsilon)$ is
the Fermi distribution function. Finnaly, the differential
conductance $G_{AB}$ can obtained from $G_{AB}=d J_{AB}/d
V_{bias}$.

Based on Eq.(10), we show the numerical results of the current
$J_{AB}$ flowing through the open AB ring in Fig.4. Fig. 4(a) and
Fig. 4(b) correspond to small and large bias voltage cases,
respectively. In both cases, the phonon-assisted sub-peaks can be
seen on the right hand side of the main peak, and the sub-peak
height increases with increasing e-ph coupling strength $\lambda$.
These results of current are similar to those in the previous
paper.\cite{N. S. Wingreen}

Next, we focus on the differential conductance $G_{AB}$
($G_{AB}=dJ_{AB}/dV_{bias}$) and its dependence on the
renormalized level $\overline{\varepsilon}_0$ or on the magnetic
flux $\phi$. In fact, $\overline{\varepsilon}_0$ and $\phi$ can be
well controlled and are continuously tunable in an experiment. The
differential conductance $G_{AB}$ is always a periodic function of
the magnetic flux $\phi$ with a period of $2\pi$. Figs. 5(a) and
5(b) show the linear conductance $G_{AB}$ at zero bias voltage.
Here a series of phonon-assisted sub-peaks exhibits in the curve
of $G_{AB}$ versus $\overline{\varepsilon}_0$, similar to that in
the transmission probability $|t_0|^2$ [see Fig. 1(a)] because of
small value of $t_{ref}$. Besides, the phonon-assisted tunneling
processes can also be observed from the amplitude of the  $G_{AB}$
oscillation versus the magnetic flux $\phi$ [see Fig. 5(b)]. While
$\overline{\varepsilon}_0 =0$ or $-1$ (i.e. at the main peak or
the 1-st sub-peak), the AB oscillation amplitudes are quite large
since the phonon-assisted elastic tunneling processes play a role
here. But at the position between two adjacent peaks (e.g.
$\overline{\varepsilon}_0 =-0.5$ or $-1.5$), the AB oscillation
amplitude is quite weak. When a small bias voltage is applied
between the left and right leads, all peaks in the curve of
$G_{AB}$-$\overline{\varepsilon}_0$, including the main peak and
phonon-assisted sub-peaks, split into two [see Fig. 5(c)], and
their positions are at $\overline{\varepsilon}_0 =m\omega_0 \pm
V_{bias}/2$. The reason is that at these values of
$\overline{\varepsilon}_0 =m\omega_0 \pm V_{bias}/2$, the
renormalized level $\overline{\varepsilon}_0$ is in line with the
left or the right chemical potential $\mu_{L,R}=\pm V_{bias}/2$,
or the distance between $\overline{\varepsilon}_0$ and $\mu_{L,R}$
is just $m\omega_0$. The behavior of the conductance $G_{AB}$
versus the magnetic flux $\phi$ for small bias is similar with
that of the linear conductance [see Figs. 5(b) and 5(d)]. Note
that at zero bias or at small bias ($V_{bias}<\omega_0$) all
tunnelings through the QD are completely coherent, and the small
amplitude oscillation in $G_{AB}$ is due to the small transmission
probability $|t_0|^2$. Finally, we investigate the large bias case
($V_{bias}>\omega_0$). At large bias $V_{bias}$ the peaks in the
curve of $G_{AB}$ versus $\overline{\varepsilon}_0$ clearly split
into two with an interval of $V_{bias}$. Moreover, some extra
sub-peaks emerge even on the left of the main peak. For example,
the sub-peak marked by "B" in Fig. 5(e) stands on the left of the
main peak at $\overline{\varepsilon}_0 = -V_{bias}/2$, and their
interval is $\omega_0$. In fact, this peak is from the inelastic
tunneling process $t_1(\varepsilon)$ and a phonon is left in the
QD with an electron tunneling through the dot. Fig. 5(f) shows the
conductance $G_{AB}$ versus the magnetic flux $\phi$ while the
level $\overline{\varepsilon}_0$ is fixed on the peak positions of
Fig. 5(e). The amplitude of the AB oscillation of the peak "B" is
very weak, but the amplitudes are quite large for other three
peaks. This gives a proof that the peak "B" is indeed from the
inelastic tunneling process and the corresponding tunneling
electron losses its phase coherent.

Since the differential conductance $G_{AB}$ is a periodic function
of the magnetic flux $\phi$, one can make the Fourier expansion:
$G_{AB}(\phi)=G^0_{AB}+G'_{AB}cos(\phi+\theta_0)$. Here the
initial phase $\theta_0$ is a direct experimentally measurable
quantity. Let us compare the measured phase $\theta_0$ with the
transmission phase $\theta$. Fig. 6 shows the phase $\theta_0$
versus $\overline{\varepsilon}_0$ for different bias $V_{bias}$.
While at zero bias voltage the characteristics of the phase
$\theta_0$, including the phonon-induced dips, are completely the
same as the transmission phase $\theta$ (see Fig. 1). When a small
bias voltage (e.g. $V_{bias}=0.2<\omega_0$) is applied between the
two leads, the phase $\theta_0$ changes slightly but can still
reflect the transmission phase $\theta$ quantitatively. So at zero
or small bias the transmission phase $\theta$, including the
intriguing characteristics due to the e-ph interaction, can be
directly observed through the measurement of differential
conductance versus the intra-dot level. For a molecular QD device,
the phonon frequency $\omega_0$ is usually from 5mev to
35mev,\cite{H. Park,B. J. Leroy} so the condition
$V_{bias}<\omega_0$ is easily reachable. On the other hand, at
large bias case (e.g. $V_{bias}=1.5\omega_0$), the phase
$\theta_0$ deviates severally from the transmission phase
$\theta$.

\section{CONCLUSIONS}

In summary, we study the influence of the electron-phonon (e-ph)
interaction on the transmission phase and the dephasing while
electron tunneling through a molecular quantum dot. It is found
that the transmission phase versus the intra-dot level exhibits a
non-monotonic behavior, and a pronounced dip emerges when the
renormalized level locates at the position of the phonon-assisted
sub-peaks. In particular, phonon-induced dips in the transmission
phase are much more apparent than the phonon-assisted sub-peaks in
the conductance. Besides, phonon-induce dephasing increases
monotonically with the e-ph interaction strength $\lambda$. The
dephasing probability $T_d$ is proportional to $\lambda^2$ at
small $\lambda$, but $T_d\propto \lambda$ for large $\lambda$. In
addition, the open AB ring device is investigated. At zero bias or
small bias, the measurement phase from the differential
conductance versus the magnetic flux is found to have the same
characteristics with the transmission phase, including the
phonon-induced dips.

\section*{ACKNOWLEDGMENTS}
We are grateful to Jianing Zhuang for his helpful discussions. This
work is supported by US-DOE under Grant No. DE-FG02-04ER46124, NSF
under CCF-052473, and NSF-China under Grant Nos. 10474125 and
10525418.

\section*{APPENDIX}
In this appendix, we present a detailed derivation for the
transmission amplitude $t_m(\varepsilon_i)$, Eq. (6). Because those
states are normalized according to $\langle
{\varepsilon,n,\alpha}\mid {\varepsilon',n',\alpha
'}\rangle=\delta_{n,n'}\delta_{\alpha,\alpha'}\delta(\varepsilon-\varepsilon')$,
only the last term in the S-matrix [Eq. (4)] contributes to the
scattering matrix element
$t_m(\varepsilon_f,\varepsilon_i)$.~\cite{N. S. Wingreen} Therefore
the scattering matrix element $t_m(\varepsilon_f,\varepsilon_i)$ is
reduced into:
\begin{eqnarray}
&&t_m(\varepsilon_f,\varepsilon_i)=\langle {\varepsilon_{f},m,R}\mid S\mid {\varepsilon_{i},0,L}\rangle\nonumber\\
&=&-\theta(t_2-t_1)\int^{\infty}_{-\infty}\int^{\infty}_{-\infty}dt_1
dt_2 e^{-\eta (|t_1|+|t_2|)}\nonumber\\
 &&\times e^{i(\varepsilon_f
+m\omega_0)t_2-i\varepsilon_it_1}\langle {\varepsilon_{f},m,R}\mid
H_1 e^{-iH(t_2-t_1)}H_1\mid {\varepsilon_{i},0,L}\rangle\nonumber\\.
\end{eqnarray}
To take the change of variables: $t_1=t_1$ and $t=t_2-t_1$, the
integration over $t_1$ now yields a $\delta$ function of energies as
$\eta\rightarrow 0^+$, and $t_m(\varepsilon_f,\varepsilon_i)$
changes into:
\begin{eqnarray}
&&t_m(\varepsilon_f,\varepsilon_i)=\langle {\varepsilon_{f},m,R}\mid S\mid {\varepsilon_{i},0,L}\rangle\nonumber\\
&=&-2\pi\delta(\varepsilon_i-\varepsilon_f-m\omega_0)\theta(t)\int^{\infty}_{-\infty}
dt e^{i\varepsilon_it} \nonumber\\
&&\times\langle {\varepsilon_{f},m,R}\mid H_1
e^{-iHt}H_1\mid {\varepsilon_{i},0,L}\rangle\nonumber\\
&=&-2\pi\delta(\varepsilon_i-\varepsilon_f-m\omega_0)\theta(t)\int^{\infty}_{-\infty}
dt e^{i\varepsilon_it}\sum_{k'}\sum_{k} \nonumber\\
&&\times t_{R k'}{t^*_{L k}}\langle {\varepsilon_{f},m,R}\mid c_{R
k'}^\dagger d e^{-iHt}d^\dagger c_{L k}\mid
{\varepsilon_{i},0,L}\rangle\nonumber\\
&=&-2\pi\delta(\varepsilon_i-\varepsilon_f-m\omega_0)\theta(t)\int^{\infty}_{-\infty}
dt e^{i\varepsilon_it} \nonumber\\
&&\times t_{R}(\varepsilon_f){t_{L}^{*}(\varepsilon_i)}\langle
{m}\mid d
e^{-iHt}d^\dagger\mid {0}\rangle\nonumber\\
&=&-2\pi\delta(\varepsilon_i-\varepsilon_f-m\omega_0)\theta(t)\int^{\infty}_{-\infty}
dt e^{i\varepsilon_it} \nonumber\\
&&\times t_{R}(\varepsilon_f){t_{L}^{*}(\varepsilon_i)}\langle
{0}\mid \frac{b^m}{\sqrt{m!}}d
e^{-iHt}d^\dagger\mid {0}\rangle\nonumber\\
&=&-2\pi\delta(\varepsilon_i-\varepsilon_f-m\omega_0)\theta(t)\int^{\infty}_{-\infty}
dt e^{i\varepsilon_it} \nonumber\\
&&\times t_{R}(\varepsilon_f){t_{L}^{*}(\varepsilon_i)}\langle
{0}\mid \frac{b^m(t)}{\sqrt{m!}}d(t)d^\dagger\mid {0}\rangle,
\end{eqnarray}
where
$|t_{L(R)}(\varepsilon)|^2=\sum_k|t_{Lk(Rk)}|^2\delta(\varepsilon-\varepsilon_{Lk(Rk)})$.~\cite{addref2,N.
S. Wingreen} At zero temperature, the above equation can be
rewritten as:
\begin{eqnarray}
t_m(\varepsilon_f,\varepsilon_i)
&=&-2\pi\delta(\varepsilon_i-\varepsilon_f-m\omega_0)\frac{t_{R}(\varepsilon_f){t_{L}^{*}(\varepsilon_i)}}{\sqrt{m!}} \nonumber\\
&&\times \theta(t)\int^{\infty}_{-\infty} dt e^{i\varepsilon_it}
Tr\{b^m(t)d(t)d^\dagger\}.
\end{eqnarray}
In order to calculate $ Tr\{b^m(t)d(t)d^\dagger\}$, we apply a
canonical transformation with:~\cite{G. D. Mahan} $
\overline{H}=e^{s}H e^{-s}$ and $s=(\lambda /\omega_0) (b^\dagger-
b){d^\dagger d}, $. Under this canonical transformation ,
Hamiltonian (1) becomes:
\begin{eqnarray}
\overline{H}=\overline{H}_{el}+\overline{H}_{ph}
\end{eqnarray}
where
\begin{eqnarray}
\overline{H}_{el}&= &\sum_{\alpha,k}{\varepsilon_{\alpha k}c_{\alpha
k}^\dagger c_{\alpha k}}+\overline{\varepsilon}_0d^\dagger
d+\sum_{\alpha,k}{[\overline{t}_{\alpha k} c_{\alpha k}^\dagger
d+H.c]} \\
\overline{H}_{ph}& =& \omega_0b^\dagger b,
\end{eqnarray}
where $\overline{\varepsilon}_0=\varepsilon_0-g\omega_0$ is the
renormalized level of the QD and $\overline{t}_{\alpha k}=t_{\alpha
k}X$, with $g\equiv (\lambda/\omega_0)^2$ and $X\equiv
\exp{[-(\lambda/\omega_0)(b^\dagger+b)]}$. Next we employ the same
approximation as one in the paper (~\cite{J. X. Zhu}),
$\overline{t}_{\alpha k}\thickapprox t_{\alpha k}$. Under this
approximation, e-ph interaction can be decoupled and
$t_m(\varepsilon_f,\varepsilon_i)$ in Eq. (13) varies into:
\begin{eqnarray}
t_m(\varepsilon_f,\varepsilon_i)
&=&-2\pi\theta(t)\delta(\varepsilon_i-\varepsilon_f-m\omega_0)
\frac{t_{R}(\varepsilon_f){t_{L}^{*}(\varepsilon_i)}}{\sqrt{m!}} \nonumber\\
&&\times\int^{\infty}_{-\infty}
dt e^{i\varepsilon_it} Tr\{b^m(t)\bar{d}(t)X(t)\bar{d}^\dagger X^\dagger\}\nonumber\\
&=&-i2\pi\delta(\varepsilon_i-\varepsilon_f-m\omega_0)
\frac{t_{R}(\varepsilon_f){t_{L}^{*}(\varepsilon_i)}}{\sqrt{m!}} \nonumber\\
&&\times\int^{\infty}_{-\infty} dt e^{i\varepsilon_it} \bar{G}^r(t)
e^{-\Phi_m(t_2-t_1)},
\end{eqnarray}
where $\bar{G}^r(t)=-i\theta(t)Tr_{el}\{\bar{d}(t)\bar{d}^\dagger
\}=-i\theta(t)\langle 0\mid\bar{d}(t)\bar{d}^\dagger\mid 0\rangle$
and $e^{-\Phi_m(t_2-t_1)}=Tr_{ph}\{b^m(t)X(t) X^\dagger\}$. Using
the method of Feynman disentangling of operators,~\cite{G. D. Mahan}
$e^{-\Phi_m(t_2-t_1)}$ can be obtained:
\begin{eqnarray}
e^{-\Phi_m(t_2-t_1)}&=&Tr_{ph}\{b^m(t)X(t) X^\dagger\}\nonumber\\
&=&\langle 0 \mid b^m(t)X(t) X^\dagger\mid 0\rangle_{ph}\nonumber\\
&=&e^{-\lambda u}\langle 0 \mid b^m(t)e^{ b^{\dagger}u^*}e^{- b
u}\mid {0}\rangle_{ph}\nonumber\\
&=&e^{-\lambda u}e^{-im\omega_0t}
\langle 0 \mid b^m e^{ b^{\dagger}u^*}\mid {0}\rangle_{ph}\nonumber\\
&=&e^{-\lambda u}e^{-im\omega_0t}\frac{(u^*)^m}{m!}
\langle 0 \mid b^m  (b^{\dagger})^m\mid {0}\rangle_{ph}\nonumber\\
&=&e^{-\lambda u}e^{-im\omega_0t}(u^*)^m ,
\end{eqnarray}
where $u=\lambda[1-e^{-i\omega_0(t_2-t_1)}]$. Substituting Eq. (18)
into Eq. (17), we have:
\begin{eqnarray}
t_m(\varepsilon_f,\varepsilon_i)
&=&-i2\pi\delta(\varepsilon_i-\varepsilon_f-m\omega_0)
\frac{t_{R}(\varepsilon_f){t_{L}^{*}(\varepsilon_i)}}{\sqrt{m!}} \nonumber\\
&&\times\int^{\infty}_{-\infty} dt e^{i\varepsilon_it} \bar{G}^r(t)
e^{-\lambda u}e^{-im\omega_0t}(u^*)^m\nonumber\\
&=&-\frac{i}{\sqrt{m!}}{\Gamma}
e^{-\lambda ^2}\sum_{l=0}^ m\frac{(-1)^{m-l}m!}{l!(m-l)!}\nonumber\\
&&\times\sum^{\infty}_{n=0}\frac{\lambda^{2n+m}}{n!}\overline{G}^r(\varepsilon_i-n\omega_0-l\omega_0),
\end{eqnarray}
where $\overline{G}^r(E)$ is the Fourier transform of
$\overline{G}^r(t)$. Here we have assumed the symmetric coupling [
$t_{L}(\varepsilon_i)=t_{R}(\varepsilon_i)$] and considered the
wide-band limits case, so $\Gamma =2\pi
t_{L(R)}(\varepsilon_f){t_{L(R)}^{*}(\varepsilon_i)}$ is independent
of the energy $\varepsilon_i$ and $\varepsilon_f$. In the wide-band
limit the Green's function $\bar{G}^r(\varepsilon)$ is easily
calculated following the standard procedure,~\cite{J. X. Zhu,A. P.
Jauho,H. Huag}
\begin{eqnarray}
\bar{G}^r(\varepsilon)=\frac{1}{\varepsilon-\bar{\varepsilon}_0+i\Gamma}.
\end{eqnarray}
From $t_m(\varepsilon_f,\varepsilon_i)$, the transmission amplitude
$t_m(\varepsilon_i)$ can be obtained: $t_m(\varepsilon_i)= \int d
\varepsilon_f t_m(\varepsilon_f,\varepsilon_i)$, and the result is
given in Eq. (6) in the text.

\newpage

\begin{figure}
\caption{(Color online) The transmission probability $|t_0|^2$ (a)
and the transmission phase $\theta$ (b) vs. the renormalized level
$\overline{\varepsilon}_0$ for the different e-ph interaction
strength $\lambda$ with $\Gamma=0.1$. The inset in (a) is the
schematic diagram for the phonon-assisted tunneling
process.}\label{fig:1}
\end{figure}

\begin{figure}
\caption{(Color online) The transmission probability $|t_0|^2$ (a)
and the transmission phase $\theta$ (b) vs. the renormalized level
$\overline{\varepsilon}_0$ for the different $\Gamma$ with the e-ph
interaction strength $\lambda=0.7$. }\label{fig:2}
\end{figure}

\begin{figure}
\caption{(Color online) (a) e-ph coupling strength $\lambda$
dependence of the dephasing probability, $T_d$. (b) The relative
dephasing probability $T_d/T_{tot}$ vs. $\lambda$. The parameter
$\Gamma=0.1$ in (a) and (b). The dot line in (b) is guide to the
eye. }\label{fig:3}
\end{figure}

\begin{figure}
\caption{(Color online) The current $J_{AB}$ vs. the renormalized
level $\overline{\varepsilon}_0$ for the different e-ph interaction
strength $\lambda$ with $\Gamma=0.1$, $\phi=0$, and $t_{ref}=0.1$. }
\label{fig:4}
\end{figure}

\begin{figure}
\caption{(Color online) (a, c, and e) are the differential
conductance $G_{AB}$ vs. the level $\overline{\varepsilon}_0$ for
the bias $V_{bias}=0$ (a), $0.2$ (c), and $1.5$ (e) at $\phi=0$. (b,
d, and f) are the differential conductance $G_{AB}$ vs. the magnetic
flux $\phi$ for the bias $V_{bias}=0$ (b), $0.2$ (d), and $1.5$ (f).
The other parameters are $\Gamma=0.1$ and $t_{ref}=0.1$. }
\label{fig:5}
\end{figure}

\begin{figure}
\caption{(Color online) The phase $\theta_0$ vs. the level
$\overline{\varepsilon}_0$ for the different bias $V_{bias}$. The
other parameters are $\lambda=0.6$, $\Gamma=0.1$, and $t_{ref}=0.1$.
}\label{fig:6}
\end{figure}

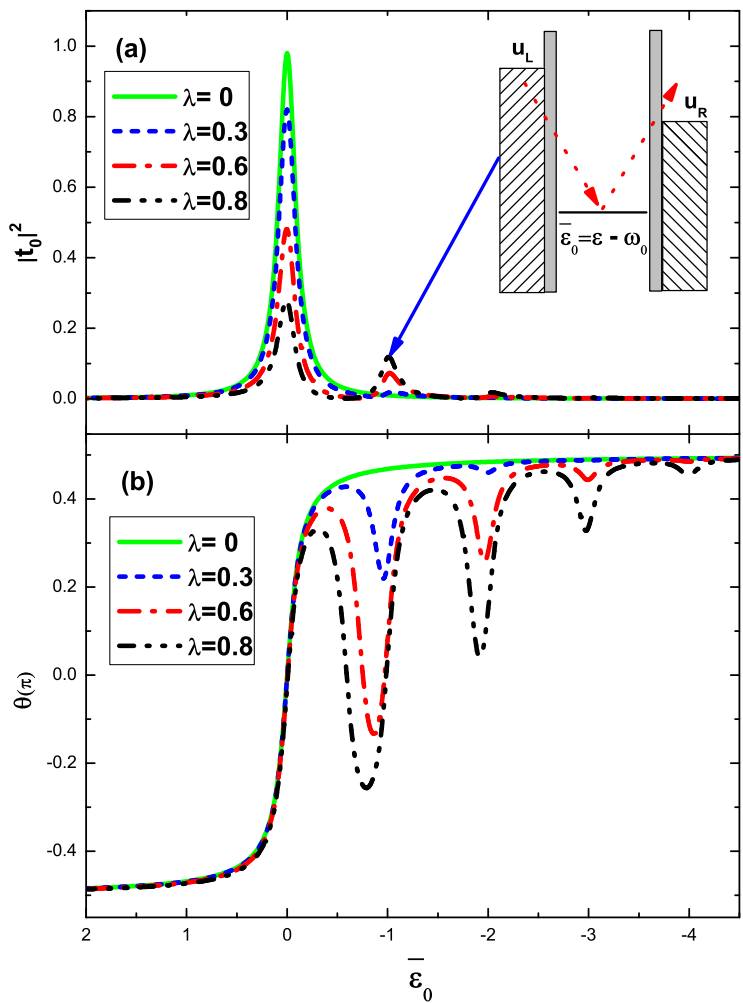
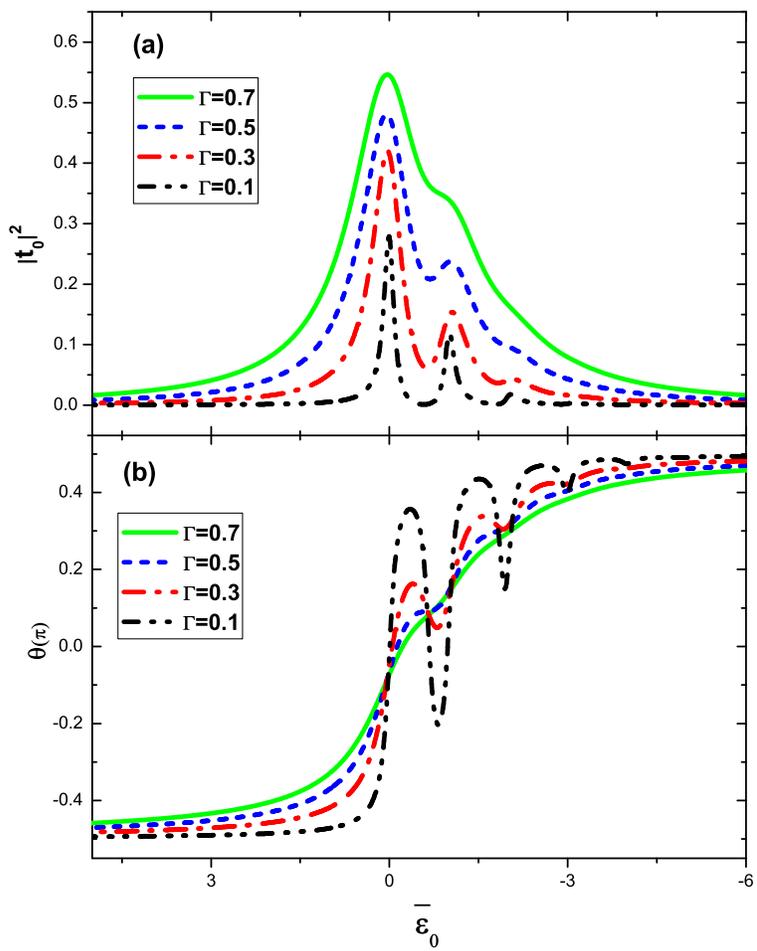
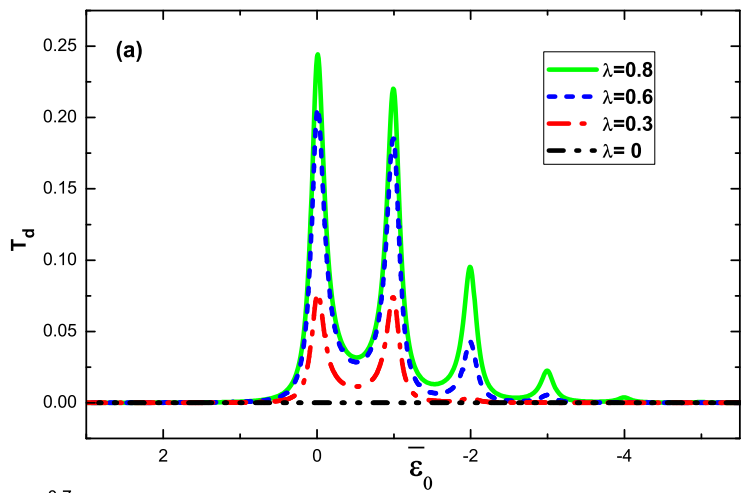
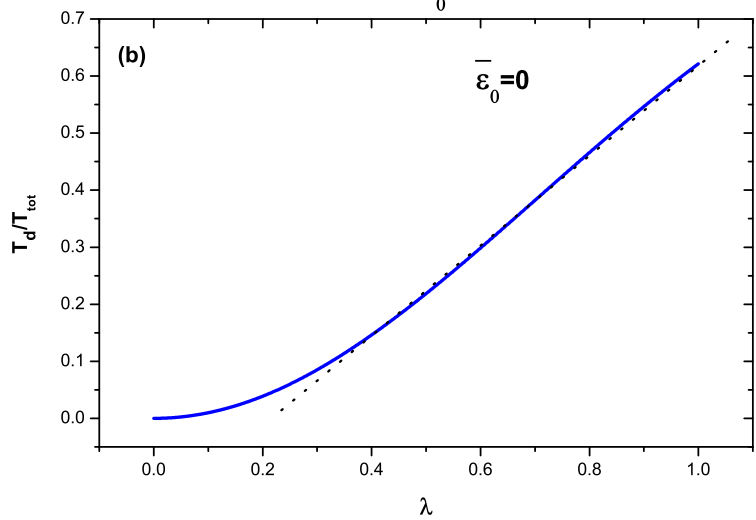
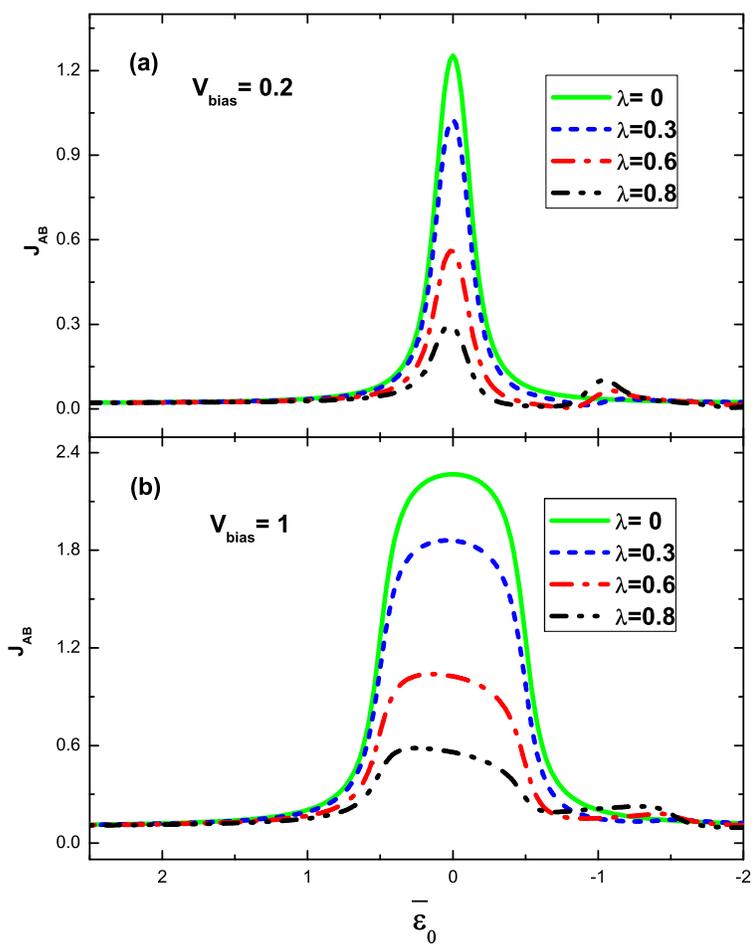
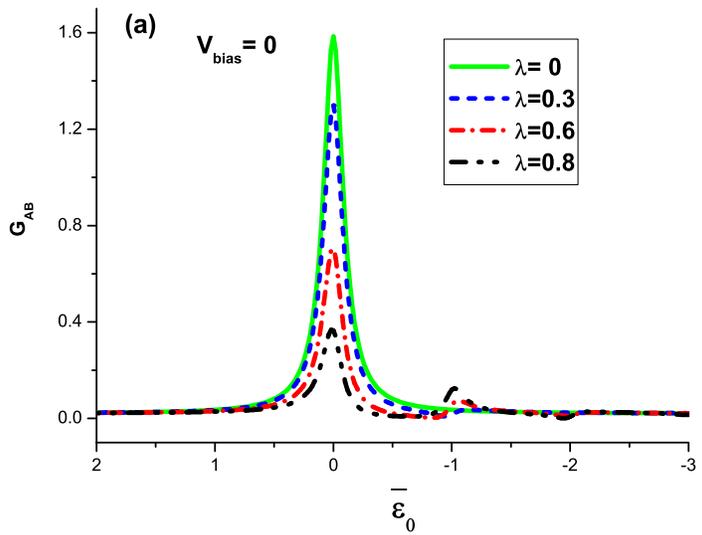
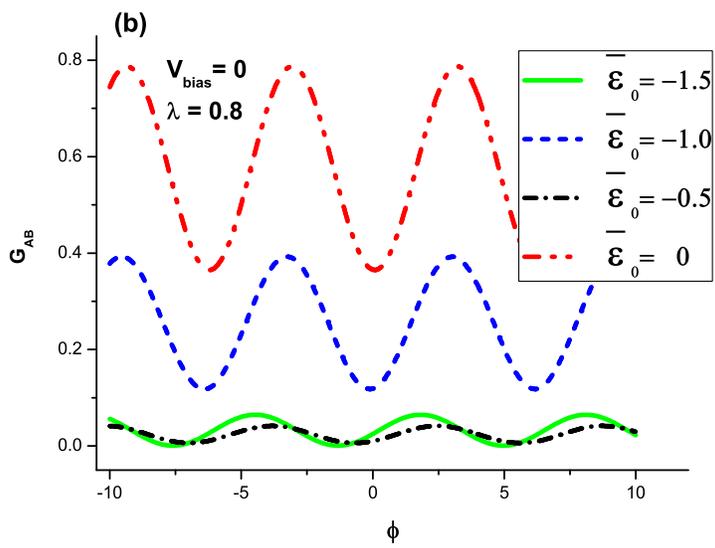
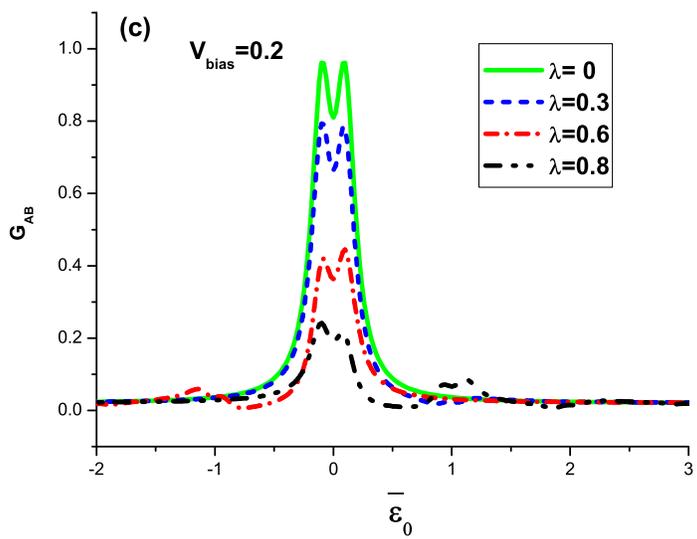
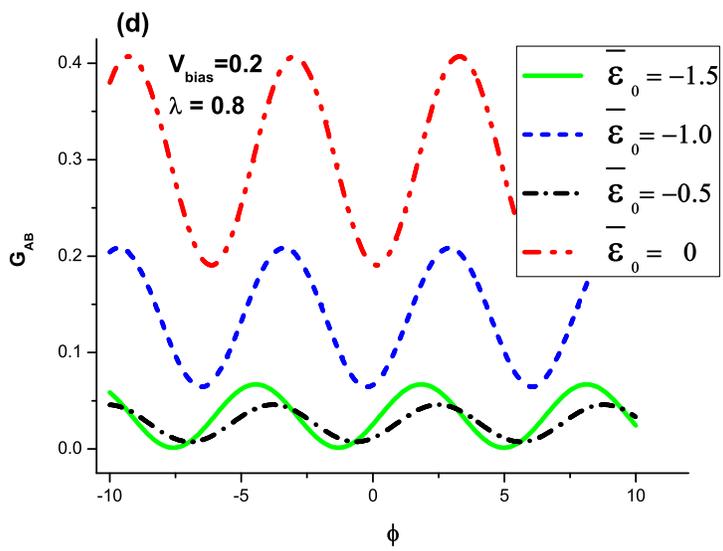
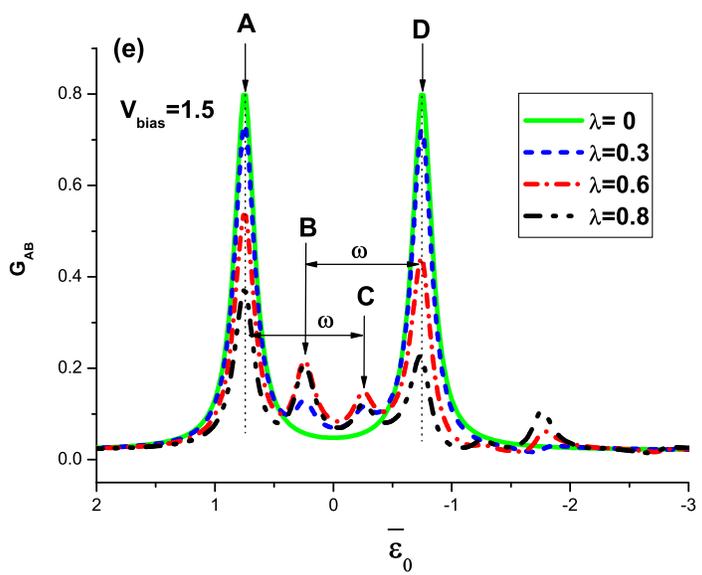
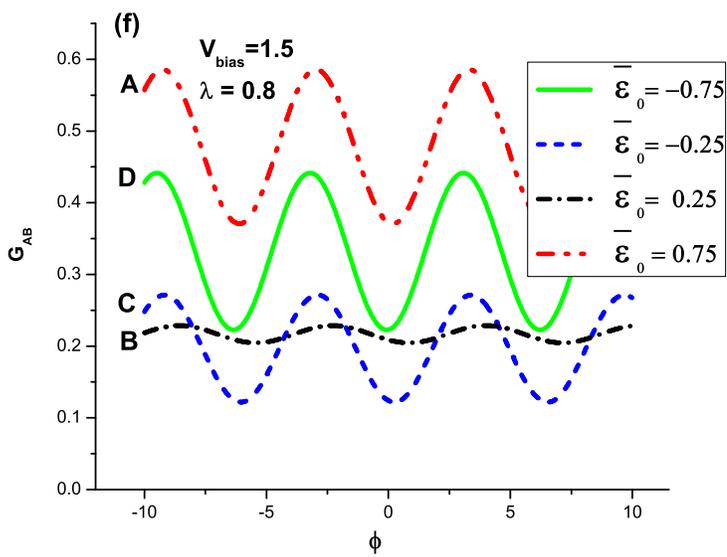
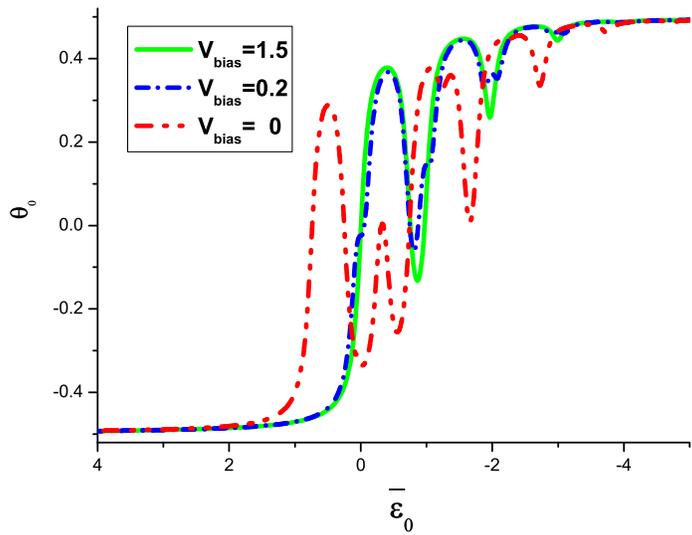
\end{document}